# A pedagogically guided LLM framework for scaffolding human–AI cocreation and preserving learners' agency and creativity

Ijaz Ul Haq
University of Lleida, Spain · ijaz.ul-haq@udl.cat

*This manuscript is currently under review.*

## Abstract

Large language models (LLMs) can generate alternatives, elaborate ideas and supply rationales, overlapping with cognitive work learners are expected to perform themselves. However, human–AI cocreation still lacks a mechanism for determining how AI can scaffold learners' creative decision-making. This study develops a cognitive–pedagogical framework for deriving the behaviour of a pedagogically constrained LLM from an educator-approved specification of each activity in a cocreative process in project-based learning. The creative-process phase supplies the activity's purpose, typical processes and pedagogical conditions; the active section of a creative technique specifies how the group works on its ideas. For each task, an educator approves four fields: a cognitive and metacognitive objective, a pedagogical objective, the task and group artefact as context, and the learner-owned product. The objective identifies the creative operations and decisions about progress, focus, criteria and next steps that remain with the group. Four translation rules derive permitted support, timing, feedback scope and prohibitions. The framework is illustrated with Viacocrea's six phases and two activities: Plan Brainstorming and a constructed Build activity using PPCo (Pluses, Potentials, Concerns, Overcoming concerns). In the proposed workflow, a controller applies the approved specification to student requests and feedback offers at approved checkpoints. Prompt engineering and approved pedagogical guidance retrieved through retrieval-augmented generation (RAG) inform a draft, which is checked for form, pedagogical function and safety before delivery. The worked examples illustrate how support can differ between activities while its constraints remain traceable to educator decisions. The framework offers educators and researchers a reusable way to derive LLM support from creative activities while preserving learners' agency and creativity.

## 1. Introduction

Creativity as a twenty-first-century skill is reflected in educational policy initiatives and curricula (OECD, 2023). Yet creativity remains a complex and multifaceted construct (Sawyer, 2021). It is shaped by social and situational contexts and develops through iterative and improvisational processes, particularly in collaborative settings (Sawyer, 2021, 2022). Such collaboration involves cognitive and social processes through which groups share and develop ideas (Sun et al., 2022). A cocreative process can organise this work into phases in which groups generate, compare, elaborate and justify ideas collectively. Each phase gives the group a purpose; the task and the section of a creative technique in use specify what learners do in a particular activity.

In this context, project-based learning (PBL) situates cocreative work in authentic, open-ended challenges that students address together, with digital tools supporting their collaboration (Haatainen & Aksela, 2021). PBL can accommodate four features Sawyer (2022) identifies in teaching the creative process: iteration, ambiguity, exploration and emergence. Groups may revisit ideas as a project changes direction, formulate a problem as well as address it, test possibilities through experimentation and develop ideas through making and doing. The cocreative process that informs this framework comprises six phases (Start, Define, Plan, Build, Conclude and Communicate), and groups use structured creative techniques across them (Pifarré, 2023; Ul Haq et al., 2024a, 2024b). In a shared digital workspace, they record short, open-ended ideas in sentence form. These written ideas offer a partial trace of the shared cognitive work through which the group's response takes shape.

A large language model (LLM) can propose ideas, connect them or decide which direction a group should take next. Its use therefore raises a central educational concern: whether it supports students' creative thinking or performs the work they are meant to practise. Rhodes's (1961) person–process–product–press distinction treats a creative product and the process behind it as different objects of study. Green et al. (2024) define creativity in terms of process, while Aru (2025) examines the importance of internal processes when assessing AI-generated output. AI-generated story ideas improved individual writers' rated products but reduced diversity across stories (Doshi & Hauser, 2024), whereas human- and ChatGPT-labelled primes had mostly null effects on divergent association scores (Medeiros et al., 2025). Neither study examined groups learning through structured creative techniques. Better group ideas alone would not show that students learned to think creatively; it would also be necessary to examine how they developed those ideas together and whether they can do similar work without AI later.

To support learning with LLMs, researchers have prompted or trained them to offer hints and guidance rather than solve students' problems outright (Kazemitabaar et al., 2024; LearnLM Team, 2024; Sonkar et al., 2024). In high-school mathematics, a less constrained AI assistant improved practice performance but impaired later unassisted performance (Bastani et al., 2025). In cocreative work, however, there is no single correct answer to withhold. A question may still draw attention to one idea, connect ideas on the group's behalf or introduce a criterion for choosing among them. Asking the model to use questions therefore does not establish that students have performed the intended creative work or directed its course. The educator must first specify which operations and decisions belong to the group in each activity; the LLM's permitted contribution can then be defined in relation to them. The unresolved question is therefore how LLM behaviour should be derived from the cognitive and pedagogical requirements of a creative process.

This study addresses that question by developing a framework for specifying the creative work of an activity before deriving LLM support. An educator draws on the current phase, the section of the creative technique in use and the task to complete a four-field activity specification: a cognitive and metacognitive objective, a pedagogical objective, the task and existing group artefact as context, and the product learners must author. This makes the educator's judgement about learners' operations and decisions explicit. The AI requirements derived from it must meet two conditions. Each permission and prohibition must be traceable to the activity specification, so its educational rationale can be examined and revised. The requirements must also prevent the model from performing creative work or directing decisions reserved for the group. The specification's adequacy thus depends on its grounding in the activity and on the limits it places on AI support. These conditions define the first research question.

> **RQ1.** *How can the cognitive, metacognitive and pedagogical objectives of an activity within a cocreative process be translated, through educator judgement, into a traceable specification of AI scaffolding that preserves learner agency?*

Building on this, the approved AI-support specification becomes operational when its requirements govern the LLM's responses. Prompt engineering expresses what the model should do and avoid, supported by examples of appropriate and inappropriate responses. Retrieval-augmented generation (RAG) supplies approved guidance for the current activity, while a controller selects the applicable specification and determines whether a student request or educator-approved feedback offer is eligible. Because these measures cannot guarantee that a draft follows the approved requirements, pre-delivery checks assess it against the specification, the work reserved for learners and safety requirements. Coordinating these components to apply the educator's decisions during interaction defines the second research question.

> **RQ2.** *How can the approved AI-support specification be implemented in a constrained LLM workflow using prompting, guidance retrieval, controlled activation and response checks for student requests and educator-approved feedback offers?*

Together, the two research questions lead to three contributions. First, the paper provides a human-in-the-loop procedure for specifying the cognitive, metacognitive and pedagogical demands of a cocreative activity and deriving permitted and prohibited LLM support. Second, it sets out a constrained workflow in which prompting, retrieved guidance, activation decisions and pre-delivery checks apply the approved specification to student requests and feedback offers. Third, it provides a reusable template for adapting the procedure to other creative processes, phases and techniques. Together, these elements make the connection between a cocreative activity's educational design and the LLM's permitted contribution explicit and open to examination.

# 2. Theoretical and technical foundations

This section brings together two foundations for the framework: creativity, cognition and learning research on the work groups undertake across phases and techniques, followed by educational AI research on LLM support, prompting, retrieval and response checks.

## 2.1 Creative process, phases and techniques in creativity research

Creativity is not a single entity; cognitive psychology links it to activities and processes through which creative work develops. The interlinked thoughts and actions involved in generating, examining and revising ideas form a creative process that can nurture creative capacities in individuals and groups. In creativity research, creative-process models differ in their phases and purposes. Wallas (1926) described preparation, incubation, illumination and verification; Amabile (1983) distinguished task presentation, preparation, response generation, validation and outcome. Creative Problem Solving includes understanding a challenge, generating ideas and preparing for action (Isaksen & Treffinger, 2004), while Sawyer (2012) synthesised earlier accounts in an eight-stage model. Creative work may also return to an earlier problem or idea (Botella et al., 2018; Lubart, 2001; Sawyer, 2021). A named phase therefore indicates a purpose and plausible kinds of work without establishing that every group is performing one cognitive operation at that moment.

Research on creative cognition identifies the kinds of work that may occur within and across phases. Problem construction, information search, idea generation, conceptual combination, evaluation and implementation can affect creative outcomes (Mumford et al., 1991). Finke et al. (1992) describe generative and exploratory processes through which people produce and interpret preinventive forms. Problem finding can alter the possibilities available for later solution development (Getzels & Csikszentmihalyi, 1976; Reiter-Palmon et al., 1997). These processes help interpret an activity but cannot be inferred from a phase name alone. The Viacocrea process moves from forming a group and defining a challenge (Start, Define), through planning and developing ideas (Plan, Build), to selecting and communicating conclusions (Conclude, Communicate). These phases describe educational purposes that groups may revisit as the project develops (Pifarré, 2023; Viacocrea, n.d.).

At the level of an activity, a creative technique gives participants instructions for working on an idea or artefact (G. F. Smith, 1998). Brainstorming, for example, asks participants to produce possibilities, build on one another's suggestions and defer criticism during generation (Osborn, 1953). Evaluation remains necessary, but its timing differs from that of idea generation.

Creative techniques can also organise the development of a provisional idea. The Creative Problem Solving PPCo tool asks participants to identify an idea's Pluses, explore its Potentials, phrase Concerns as questions and generate ways of Overcoming concerns (Creative Education Foundation, 2015, p. 45; Puccio et al., 2011). Its sections involve different forms of appraisal and generation within one technique. Brainstorming and PPCo show why a phase name alone cannot determine what a group is doing with its current idea: the technique's instructions and the section in use also matter.

## 2.2 Group creativity, scaffolding and creative metacognition

A group's creative product can be assessed for originality and effectiveness (Runco & Jaeger, 2012); however, it does not disclose how members arrived at it. Green et al. (2024) define creativity in terms of internally directed attention constrained by a generative goal; Vallée-Tourangeau (2026) argues that making and engagement with external materials can themselves transform thought. Aru (2025) cautions against treating similar human and AI outputs as evidence of similar internal processes. In a shared workspace, an entry records a contribution to the emerging artefact but cannot by itself show who framed the problem, evaluated an option or changed the group's direction. Creative activity unfolds as members build on, combine and appraise one another's ideas while working with the shared artefact (Sawyer, 2022; Sawyer & DeZutter, 2009).

In such exchanges, the group's field of possible ideas can change when a member or an AI system highlights one fragment, proposes a category or introduces an example. In group brainstorming, production blocking can restrict contributions, while exposure to others' ideas may stimulate generation or narrow its variety (Diehl & Stroebe, 1987; Dugosh et al., 2000; Nijstad & Stroebe, 2006; Paulus & Brown, 2007; S. M. Smith et al., 1993). Written contributions can alter participation conditions, although their effects in this setting have not been tested (Paulus & Yang, 2000). In B. C. Lee and Chung's (2024) experiments, ChatGPT improved ratings of individuals' ideas, whereas Meincke et al.'s (2025) reanalysis reported less diversity across the pooled ideas. Anderson et al. (2024) similarly reported ideas that were less distinct and weaker responsibility for them under LLM assistance. Their different settings and outcomes call for examining how and when external cues affect a group's search.

Accordingly, scaffolding research distinguishes assistance that structures learners' activity from assistance that problematizes it, prompting them to examine their work more deeply (Reiser, 2004). Support is contingent on learners' activity and may be reduced as competence develops (van de Pol et al., 2010). Constructive alignment relates intended learning to the tasks and support through which students practise it (Biggs, 1996). These accounts provide a pedagogical reason to ask what learners are expected to construct and how assistance can support their participation. Applied to cocreative work, these accounts direct attention to what feedback prompts groups to do with their ideas and what it does for them.

Feedback is one form of such assistance: it can orient learners to a goal, their current progress and possible next steps (Hattie & Timperley, 2007). The Interactive–Constructive–Active–Passive (ICAP) framework distinguishes forms of engagement, including learners constructing ideas and building on one another's contributions (Chi & Wylie, 2014). In collaborative creativity, assistance that supplies an idea may improve a product while removing an opportunity for learners to construct or negotiate it.

Alongside generating and appraising ideas, creative metacognition concerns how people understand, monitor and control their creative activity (Lebuda & Benedek, 2023). Monitoring concerns the group's progress toward its goal; control concerns decisions about what to attend to, which strategy or criterion to use and when to move on. In a group, such regulation can be negotiated among members rather than inferred from any individual's short entry. Järvelä and Hadwin (2013) distinguish another participant's prompts to regulation from socially shared regulation by the group. This distinction matters for AI support: an external cue may help members notice a feature of their work, while a directive about which idea to pursue can settle a decision they might otherwise negotiate.

Earlier studies using Viacocrea measured the novelty and flexibility of group-authored ideas and examined real-time novelty feedback (Ul Haq & Pifarré, 2026; Ul Haq et al., 2024a, 2024b). Such measures can describe features of entries, but a score alone does not show whether students have practised generation, comparison or revision. They cannot determine who should generate or appraise ideas, or who should decide when to change the group's direction.

## 2.3 LLMs and the need for pedagogical constraint

LLMs can generate alternatives, elaborate and compare ideas, offer reasons and suggest a group's next direction. In cocreative learning, these operations and decisions may be assigned to students. Without knowing what a particular activity expects students to do, a general-purpose LLM can give a useful answer that also performs their creative work for them. External aids can shift cognitive demands away from the person performing a task (Risko & Gilbert, 2016). Studies in mathematics and creative tasks distinguish performance with AI assistance from later unaided performance (Bastani et al., 2025; Kumar et al., 2025). In AI-assisted writing, individual stories received higher creativity ratings while diversity across stories declined (Doshi & Hauser, 2024). These findings make the students' own participation in the creative process an essential consideration when designing LLM support.

Educational LLM systems have been designed to offer hints, questions or conceptual guidance instead of full solutions (Kazemitabaar et al., 2024; Liffiton et al., 2023; Sonkar et al., 2024). A tutor addressing a determinate problem can identify the answer it should withhold. In a creative activity, students may be expected to propose alternatives, connect ideas, compare them or decide what matters. A question can perform some of that work: "What if you designed a phone app?" introduces an option, while "Could you develop the second idea?" chooses a focus. Conversely, a factual definition may be appropriate when students are meant to apply the concept it explains. Pedagogical guidance must therefore identify the work assigned to the group and assess an LLM reply by the contribution it makes to that work.

That guidance must also address when the LLM intervenes and who controls the next move. Mixed-initiative research examines the timing and interruption cost of unsolicited help (Horvitz, 1999). The Co-Creative Framework for Interaction Design describes initiative and turn-taking in human–AI creative systems (Rezwana & Maher, 2023), while Lubart (2005) distinguishes roles such as coach and colleague. Human–automation research separates information acquisition, analysis, decision selection and action implementation as functions that may be allocated differently (Parasuraman et al., 2000). For cocreative learning, these distinctions require educational decisions about the content and timing of assistance and the authority the group retains. A pedagogically constrained LLM must follow those decisions.

## 2.4 Prompting, retrieval and response checks

Prompt engineering expresses an LLM's task, permitted behaviour and response form through instructions and examples (White et al., 2023). In cocreative learning, these instructions can specify the kind of cue the model may give while leaving idea generation, appraisal or selection to the group.

A prompt can state that boundary, but a fluent response may still cross it by supplying an idea or directing the group's attention. The model's instructions must therefore be paired with guidance about the activity and checks on its replies.

Retrieval-augmented generation (RAG) supplies external material to inform a response (Lewis et al., 2020). It can provide approved pedagogical guidance for the current phase and creative technique section, along with a factual source when the task calls for one. Retrieval does not determine whether help is appropriate at that moment or give retrieved text authority over the educator's instructions. Student entries and source passages may contain instructions that redirect the model (Greshake et al., 2023). The system must treat them as information to use or assess, rather than as instructions that change the approved activity.

Response checks assess a draft before it reaches students. Programmable guardrails can check required fields and allowed response forms (Rebedea et al., 2023), while RAG evaluations can assess whether an answer is supported by retrieved material (Es et al., 2024). These checks cannot determine from form or source support alone whether a question has generated an idea or chosen the group's focus. An LLM-based judge may help assess that function in context, but its judgements can be biased or mistaken (Zheng et al., 2023). Its reliability must be tested against human-labelled examples of support that preserves or takes over the group's assigned work.

## 2.5 Contemporary work on creativity and AI

Recent work addresses human authorship, creative process, AI roles and guarded educational support from different starting points. Table 1 shows what each approach contributes and the question it raises for a group working within a particular creative technique.

**Table 1.** Contemporary research and its relationship to the present design problem.

| Work | Contribution relevant to this paper | Question left open for cocreative learning |
|---|---|---|
| **Sherson et al. (2026)** | FERC organises iterative human–AI work around framing, exploration, refinement and commitment. | How should a classroom activity assign idea production, selection and direction between a group and an LLM within a specific technique? |
| **Rafner (2026)** | CoCREATE-AI provides guidance for reporting human–AI interventions. | What prospective rules should govern the intervention being reported? |
| **Cho (2026)** | Distinguishes forms of AI support and substitution in generative learning in mathematics. | How are those roles specified for successive sections of an open-ended group creativity technique? |
| **Vallée-Tourangeau (2026)** | Emphasises making and external representations in the creative process. | How can AI support interaction with a shared artefact while leaving its development to students? |
| **Aru (2025)** | Separates creative products from the human processes behind them. | What evidence of learner-authored work is needed alongside a group product? |
| **Medeiros et al. (2025)** | Reports mostly null effects of human- and ChatGPT-labelled primes on divergent association scores. | Under which activities and forms of help does AI affect a group's creative work? |
| **Green et al. (2024)** | Gives a process account centred on internally directed attention and a generative goal. | How should internal attention and group interaction be considered together in educational support? |
| **Vinchon et al. (2023)** | Surveys opportunities and risks across forms of human–AI creative collaboration. | Which instructional role should an LLM occupy in a given cocreative activity? |
| **Liffiton et al. (2023)** | Demonstrates guarded LLM assistance in programming education. | How should assistance be constrained when the student-owned work has no single correct solution? |

Across these approaches, creative work, AI roles and intervention design are described at different levels. For this paper, the unit that needs specification is the group's current activity. Its phase gives the purpose and typical work, the creative technique section specifies what learners do with their ideas, and the task provides the context. The framework uses these together to determine what work remains with the group and what assistance the LLM may give.

# 3. The cognitive–pedagogical framework

The cognitive–pedagogical framework has two connected parts (Figure 1). For RQ1, an educator combines the creative-process phase, the active section or step of a creative technique and the task to specify the work learners must perform and the support an LLM may provide. The specification sets two boundaries: the LLM must not perform learners' cognitive work (functional substitution) or take control of the group's focus, criteria or next step (attentional substitution). Section 3 explains how these boundaries are derived and expressed. For RQ2, Section 4 describes how a constrained workflow applies the approved specification to student requests and educator-approved feedback offers through prompting, retrieved guidance and checks before delivery.

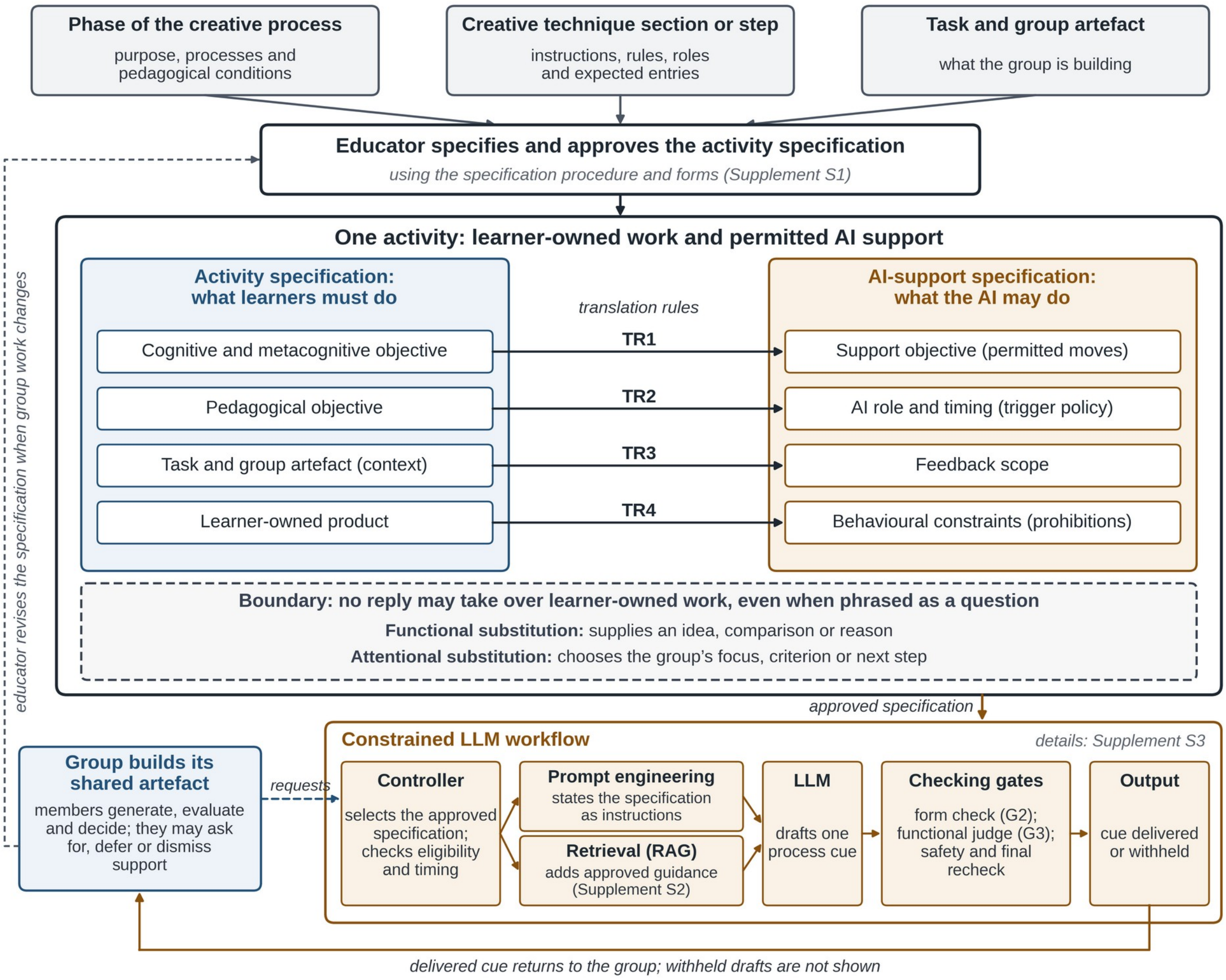

**Figure 1.** The cognitive–pedagogical framework. The educator uses the creative process, its phase, creative technique section and task to specify learner-owned work and permitted AI support (RQ1). A controller applies that specification through prompting, retrieved guidance and pre-delivery checks (RQ2).

## 3.1 From phase and creative technique to an activity specification

For each of Viacocrea's six phases, this study identified its stated purpose, what it asks groups to do and what they work on. These descriptions were then interpreted in light of accounts of creative cognition (Finke et al., 1992; Mumford et al., 1991; Puccio et al., 2011). Table 2 presents plausible cognitive processes and pedagogical conditions for each phase. These interpretations guide the educator, who also considers the task and active creative technique section when assigning work to learners.

To carry these phase interpretations into an activity specification, the framework uses labels for possible cognitive processes, such as FRAME, GENERATE, EVALUATE and SELECT. These are terms proposed for the framework, not prescribed by Viacocrea; researchers may add a label when its basis is documented. Here, EVALUATE and SELECT describe appraisal and choice among ideas. The specification also records how the group monitors progress and who decides which criteria to use, where to focus and when to move on (Lebuda & Benedek, 2023).

**Table 2.** Research-informed interpretation of Viacocrea's six phases for the proposed framework.

| Creative-process phase and stated purpose | Plausible cognitive processes | Phase-level pedagogical conditions | Research informing interpretation |
|---|---|---|---|
| **Start:** form a working group and explore the challenge. | Gather relevant knowledge and begin constructing the problem (GATHER, FRAME). | Establish shared task understanding and participation norms so members' knowledge is available to the group. | Mumford et al. (1991); Paulus and Brown (2007); Wallas (1926). |
| **Define:** identify, specify and understand the challenge. | Construct and appraise problem statements (FRAME, EVALUATE). | Keep problem framing open to revision as members learn more about the challenge. | Getzels and Csikszentmihalyi (1976); Reiter-Palmon et al. (1997); Reiter-Palmon and Robinson (2009). |
| **Plan:** identify main elements and plan actions. | Generate and develop options, then compare or choose possible actions when the activity calls for it (GENERATE, COMBINE, ELABORATE, EVALUATE, SELECT). | Keep proposed actions connected to the challenge and available for group examination before committing to a plan. | Guilford (1967); Mumford et al. (1991); Puccio et al. (2011). |
| **Build:** organise and represent ideas, generating alternatives and patterns. | Combine, transform, elaborate and generate ideas through work on representations (COMBINE, TRANSFORM, ELABORATE, GENERATE). | Keep shared representations accessible to members for questioning and revision. | Finke et al. (1992); Glăveanu (2014); Sawyer (2012). |
| **Conclude:** select conclusions that address the challenge. | Evaluate and select conclusions, and justify them (EVALUATE, SELECT, JUSTIFY). | Make criteria and reasons explicit while specifying who has authority to set them in the activity. | Cropley (2006); Mumford et al. (1991); Rietzschel et al. (2006, 2010). |
| **Communicate:** present and make a case for the results. | Shape and justify an audience-sensitive account (COMMUNICATE, JUSTIFY). | Preserve group authorship of claims and the choice of what to emphasise. | Isaksen and Treffinger (2004); Puccio et al. (2011); Sawyer (2012). |

*Note.* Phase purposes paraphrase Viacocrea (n.d.); processes and conditions are proposed interpretations. Brainstorming's deferral of criticism and PPCo's appraisal apply to their respective techniques, not entire phases.

The phase interpretation alone does not specify an activity. A researcher applying the framework to another creative process first documents its phase descriptions and the instructions of its techniques. For each technique, the reusable template records its source, sections or steps, targeted work, rules, roles, sequence and expected entries. A step receives a separate specification when it changes the work assigned to learners; a technique without named divisions is treated as one section. The educator then combines the phase, active section or step, and current task to specify the activity. Table 3 shows what each level contributes and what the educator must decide. Supplement S1 provides the forms and worked records.

**Table 3.** Sources and decisions used to specify one cocreative activity.

| Level | What it contributes | What the educator decides | Result for the activity |
|---|---|---|---|
| **Creative process** | A sequence of phases and the possibility of revisiting earlier work. | Which process applies and who controls transitions. | The phase currently in use. |
| **Phase** | A stated purpose, with plausible processes and pedagogical conditions interpreted from research (Table 2). | Which of these matters for the present task. | The activity's purpose and relevant conditions. |
| **Creative technique section or step** | Instructions for working on an idea or artefact, including roles and expected entries. | Which student work and decisions the active unit calls for. | The targeted operations and expected entries. |
| **Activity** | One active section or step in one phase for one task and group artefact. | The cognitive and metacognitive objective, pedagogical objective and learner-owned product, including who controls focus and transitions. | An educator-approved activity specification. |

The distinction matters when a technique is used in more than one phase. Brainstorming may generate possible problem framings in Define or possible actions in Plan. Its instruction to generate options remains, but the object of that work and the product learners must author change (Getzels & Csikszentmihalyi, 1976; Mumford et al., 1991). A subsequent comparison or selection requires its own approved activity rather than being inferred from the phase name. The educator resolves uncertain interpretations before approving AI support.

## 3.2 From activity specification to AI-support specification

The approved activity specification defines what learners must do. Four translation rules derive the assistance an LLM may give in relation to that work (Table 4). Their outputs form the AI-support specification: a support objective, an AI role and timing, a feedback scope and behavioural constraints.

**Table 4.** Translation from activity specification to AI-support specification.

| Activity specification (input) | Rule | Translation | AI-support specification (output) |
|---|---|---|---|
| **Cognitive and metacognitive objective** | TR1 | Invite the assigned creative work and group-owned monitoring without supplying its product or making its regulatory choices. | Support objective. |
| **Pedagogical objective** | TR2 | Select permitted structuring or problematizing moves and specify when help may be requested or offered. | AI role and timing (trigger policy). |
| **Task and group artefact (context)** | TR3 | Scope feedback to the set, item, relation or rationale on which learners work; do not select a focus they own. | Feedback scope. |
| **Learner-owned product** | TR4 | Prohibit replacing what students must construct and the group-owned decisions recorded in the objective; record educator-approved exceptions. | Behavioural constraints. |

### *3.2.1 Cognitive and metacognitive objective to support objective (TR1)*

TR1 translates the activity's cognitive and metacognitive objective into a permissible form of help. In Plan Brainstorming, students generate and combine options, monitor the breadth of their ideas and decide when to end generation. The LLM may ask what different directions they see among their entries. Embedding a candidate idea in that question would contribute to generation and may shape subsequent ideas (S. M. Smith et al., 1993). Declaring that the group has produced enough ideas would take over its decision to stop.

#### 3.2.2 Pedagogical objective to AI role and timing (TR2)

TR2 uses the pedagogical objective to set the AI's role and when help may be requested or offered. In Plan Brainstorming, an objective of broad participation may permit a structuring cue asking whether every member has had a chance to contribute. In a Build activity that calls for revising a shared representation, a problematizing question may ask what remains uncertain in that representation (Reiser, 2004). The role follows the approved activity rather than its phase name. Students may request help; they may also defer or dismiss a feedback offer. Offer limits may taper across repeated activities, but fewer offers alone do not establish that the group can work independently (van de Pol et al., 2010).

#### 3.2.3 Task and group artefact to feedback scope (TR3)

TR3 uses the task and current group artefact to set the scope of feedback: a set of entries, one entry, a relation among entries or a rationale. In Plan Brainstorming, the entries form an unranked set of possibilities. An AI prompt that singles out one entry for development would choose the group's focus before it has decided to converge, an instance of attentional substitution. After the group selects an entry in a separately approved activity, a question about that entry may become permissible.

#### 3.2.4 Learner-owned product to behavioural constraints (TR4)

TR4 sets prohibitions that apply regardless of the AI's role or feedback scope. The LLM must not supply the product learners are assigned to construct or make decisions reserved for the group. In a Conclude activity assigning evaluation, selection and justification (Table 2), a model verdict or ready-made rationale would perform that work for learners. In Brainstorming, supplying a candidate idea would perform part of the generation. An educator may approve a model-generated idea for an activity in which learners critique it, or a sourced factual clarification when providing the fact falls outside their assigned work. These exceptions must be recorded in the approved specification; an uncertain response does not create a new permission.

### 3.3 Protecting learners' creative work and decisions

When approving an activity, the educator identifies the work reserved for the group in its current phase, creative technique section or step, and task. This includes the product learners must author, the work needed to produce it, and the group's monitoring and decisions about focus, criteria and when to move on. Even a question may cross these limits by supplying content or directing a choice.

Five questions guide both approval of AI moves and assessment of draft responses: (1) Would it construct content learners are assigned to create? (2) Could learners adopt its content in place of their own product? (3) Would it bypass the assigned creative work? (4) Would it declare that enough work has been done or choose the group's focus, criterion or next step? (5) Does it carry out the approved supportive move? A response may both supply content and redirect attention. The educator resolves uncertain move types before interaction; an uncertain draft is withheld before delivery.

### 3.4 Applying the framework to two creative techniques

The framework is illustrated with Brainstorming in Plan and PPCo in Build (Table 5). In Brainstorming, the group generates possible actions without ranking them. In the PPCo example, the group starts with an idea it selected earlier. It identifies the idea's strengths (Pluses), considers what it might achieve (Potentials), phrases concerns as questions (Concerns) and proposes ways to address them (Overcoming concerns) (Creative Education Foundation, 2015, p. 45; Puccio et al., 2011). The PPCo guide places the technique in its Develop stage. Here, it serves as an example in Viacocrea's Build phase; this article does not claim that Viacocrea classrooms have used it. Because each PPCo step asks the group to do different work, the educator specifies the permitted AI support for each step. In this example, the group also revises its idea in the final step.

**Table 5.** Plan Brainstorming and a Build PPCo sequence, with the AI support derived for each.

| Specification field | Plan Brainstorming | Illustrative Build activity using PPCo |
|---|---|---|
| **Phase and technique section or step** | Plan; Brainstorming generation section. | Build; PPCo's Pluses, Potentials, Concerns and Overcoming concerns, each specified separately. |
| **Cognitive and metacognitive objective** | Generate and combine entries (GENERATE, COMBINE); monitor breadth and decide when to end generation. | Pluses: appraise strengths (EVALUATE); Potentials: elaborate possible gains (ELABORATE); Concerns: appraise and frame questions and choose a priority (EVALUATE, FRAME, SELECT); Overcoming concerns: generate and combine fixes into a revised idea (GENERATE, COMBINE). The group decides which concern matters and when to change steps. |
| **Pedagogical objective** | Defer judgement, build on members' entries and maintain breadth before convergence. | Assess and strengthen the group's selected idea while retaining authority over criteria, the priority concern and the final revision. |
| **Evidence of targeted work** | Student-authored alternatives, build-on links and discussion of breadth and closure. | Group-authored strengths, possibilities, concern questions, fixes and a revised solution; discussion of criteria and priorities. |
| **Group artefact and feedback scope (TR3)** | Unranked set of short group entries; feedback addresses the set. | One group-selected idea with entries per step; feedback concerns only the active step. |
| **Learner-owned product** | Members' candidate directions and the group's decision to end generation. | Strengths, possible gains, concern questions, priority concern, candidate fixes and revised solution. |
| **Support objective (TR1)** | Invite inspection of breadth, participation or links among the group's entries without naming a candidate. | Ask the group to inspect its active-step entries and what the step still requires without supplying it. |
| **AI role (TR2)** | Structure participation and attention to the whole set. | Prompt the group to appraise and revise its own idea. |
| **Behavioural constraints (TR4)** | Do not supply ideas, rank or single out entries, or close generation. | Do not supply strengths, gains, concern questions, fixes, a priority concern or the revised solution. |
| **When AI may respond or offer help** | Student request or approved checkpoint after initial entries during generation; at most one offer per section use. | Student request or approved checkpoint after group work in the active step; at most one offer per step use. Concern questions must precede any offer on that step. |
| **When the specification changes** | Moving to evaluation selects another approved activity or requires educator approval. | Changing PPCo steps selects that step's approved specification; the LLM does not choose the next step. |

*Note.* PPCo's steps, modification and improved solution statement are described by the Creative Education Foundation (2015, p. 45). The Build placement, process labels and assignment of revision to the final step are proposed for this example.

## 4. Operationalisation in a constrained LLM workflow

Once the activity and AI-support specifications are developed in Section 3, RQ2 asks how those decisions govern a response during group work. This article proposes a controller that applies an *activity contract*: the versioned, machine-readable form of the activity and AI-support specifications, with their approved guidance and component versions. Before calling the LLM, the controller screens the input, selects the contract for the current activity, checks eligibility and retrieves matching guidance. It then checks the model's draft for form, function and safety, and confirms that the activity is still current before delivery. The LLM cannot change the contract or advance the group's phase.

### 4.1 Prompt architecture

The prompt presents the contract for the current activity. It identifies the phase and creative technique section or step, the product and decisions reserved for the group, the help allowed and the contributions forbidden. In Plan Brainstorming, it protects the generation of an unranked set of ideas. In PPCo, it follows the current step and excludes strengths, concerns or fixes the group has yet to formulate. Research on tutoring prompts indicates that instructions can be refined for a supplied instructional goal (Celik et al., 2026; U. Lee et al., 2026). In this framework, the educator approves that goal before it is expressed in a prompt.

The approved contract and vetted guidance occupy trusted system or developer instructions. Student requests, group entries and earlier dialogue are task data and cannot revise those instructions (Greshake et al., 2023; Wallace et al., 2024). Each draft declares an approved supportive move in a structured response. Later checks compare that declaration with what the response actually says. The educator resolves any conflict between a proposed move and a prohibition before approving the activity.

The controller reads the current phase and technique section or step from the workspace or educator setting, selects its approved contract and passes only the entries relevant to the response. Workspace access follows authorised group and educator roles; the generator, judge and safety screen process the text needed for their respective checks under specified collection and retention arrangements. The proposed workflow covers written entries. Drawings and images require separate specification and testing.

### 4.2 Retrieving pedagogical guidance

Retrieval supplies the pedagogical guidance associated with the current activity. Retrieval-augmented generation conditions a response on external information (Lewis et al., 2020); here that information consists of approved records for the framework’s core rules, the phase, the creative technique section and the assigned cognitive processes. The controller loads required records by approved identifier and version rather than by similarity. Vector ranking could later select among optional, approved examples, but the present draft collection contains none. Retrieved guidance informs the response without granting an AI move or providing answers to the group’s project task. Supplement S2 contains the draft guidance records.

The educator approves the contract and guidance together. Phase guidance describes purpose and pedagogical conditions, while the activity contract determines which moves the LLM may make. A response could faithfully reflect a retrieved record and still choose an idea or criterion for the group. Measures of faithfulness to retrieved material (Es et al., 2024) therefore cannot replace a check against the learner-owned product and the group’s decisions.

During PPCo’s Concerns step, the controller loads the approved Build and Concerns guidance. The group writes its own concern questions and chooses which to address. Once it has written those questions and explicitly requests approved prioritisation help, the LLM may ask which one the group wants to address first. Neither the request nor the retrieved guidance permits the model to supply or choose a concern. Moving to another PPCo step selects the contract and guidance approved for that step.

### 4.3 Responding to students and offering feedback

Student requests and feedback offers use the same activity contract, but differ in who initiates help (Table 6). A request for an idea, rationale or decision assigned to the group does not permit the LLM to provide it. If the educator has approved a boundary explanation, the response may explain the limit and invite the group to continue. An explicit fact-help button can activate a preapproved exception for information outside the assigned creative work. The response must stay within that factual scope and leave any assigned inference to learners.

An offer can interrupt a group that has not requested help. It therefore requires an approved checkpoint, sufficient group-authored work, a limit on how often offers occur and a minimum interval between them. The offer is a brief cue about the current step, which students may defer or dismiss. The controller remains silent when these conditions are not met (Horvitz, 1999; Parasuraman et al., 2000; van de Pol et al., 2010).

**Table 6.** Student requests and feedback offers under one activity contract.

| Entry route | When it applies | Permissible response | Student control |
|---|---|---|---|
| **Student asks for process help** | Current activity is known and approved. | One relevant process cue about the group's own work. | Ask a follow-up, ignore or stop. |
| **Student asks AI to do the task** | The current activity is approved; the request asks for learner-owned work. | If approved, explain the boundary and offer a process question; otherwise use an authorised fallback, abstain or refer. | Group retains the idea, concern or decision. |
| **Student asks for a fact** | Explicit fact-help button, approved factual scope and verifiable source. | Give a scoped fact with a source, or refer; do not supply the assigned inference. | Use and evaluate the information. |
| **System offers checkpoint feedback** | Approved checkpoint after enough group work and at the specified time; the group can dismiss it. | One brief invitation about the current step; a dismissed move is not reused. | Dismiss, defer or act; no automatic phase change. |

## 4.4 Checking responses before delivery

Both routes in Table 6 pass through automated checks before a response reaches students (Table 7). The educator approves the activity and its response rules in advance; the controller applies them to each request or offer. Input is screened before a model call. For use with minors, an attributed risk disclosure receives a fixed acknowledgement in that student's private channel and alerts the educator. A disclosure in a shared entry alerts the educator without replying to the group or assuming who wrote it. Institutional oversight of shared entries operates independently of LLM activation.

After the LLM drafts a response, a form check (G2) tests its declared move and required structure. A separate LLM judge (G3) assesses the draft against the work and decisions reserved for the group, treating student entries as task data (Hines et al., 2024; Wallace et al., 2024). Five of its six checks address substitution of learner-owned content, unauthorised decisions, premature closure, activity override and unapproved content. The sixth asks whether the text carries out its declared supportive move. Together, these checks apply the five questions in Section 3.3. For an approved factual exception, G3 also checks the claim against its source.

A draft that fails a check, or receives an uncertain judgement, is withheld. A limited retry may use a coarse failure category without revealing the judge's detailed reasoning. If no permissible reply remains, a student request receives an authorised fallback or educator-referral notice; a feedback offer ends silently. Output safety screening and a final check of the activity and relevant entries precede delivery. Passing these checks does not establish that a response is useful to learners.

The checks run automatically for individual drafts; routine delivery does not depend on an educator approving each response. Before classroom use, however, independent raters would need to label test responses across the intended languages, age groups, activities and two help routes. Those labels would establish how often the judge accepts a violation or rejects permissible support. An unvalidated setting remains blocked at G0, and relevant component changes require renewed testing. This matters because the judge may share the generator's errors (Rebedea et al., 2023; Zheng et al., 2023). Supplement S3 describes the proposed testing procedure.

**Table 7.** Activation and checks before delivery.

| Check | Condition checked | Action on failure |
|---|---|---|
| **Input safeguarding** | Screen text before a model call and route risk disclosures. | Give an attributed student a private fixed acknowledgement and alert the educator. For a shared disclosure, alert the educator without replying to the group. |
| **Eligibility (G−1)** | A student request or approved feedback checkpoint may run now. | No model call; an approved fallback may be available for a request, while an offer stays silent. |
| **Approval (G0)** | The contract matches the current phase, section or step, learner-owned work and group decisions; the setting and component versions have passed offline checks. | Stop; obtain approval for a changed specification or validation for an untested setting. |
| **Guidance (G1)** | Required approved guidance matches the activity. | Stop without generation. |
| **Form (G2)** | The draft uses an allowed move and the required language and structure, including an approved source identifier for a factual exception. | Withhold; a limited retry may be allowed. |
| **Function (G3)** | A separate LLM judge applies six checks against learner-owned work and group decisions, including source checking for an approved factual claim. | Withhold on a failed or uncertain result; a limited retry may be allowed. |
| **Output safeguarding** | Screen the draft before it reaches students. | Withhold unsafe output. |
| **Final recheck** | The activity, entries referred to and basis for the cue remain current. | Discard a stale draft. A request may re-enter once under current checks; if it is still stale, give the fixed referral notice. A lapsed offer stays silent. |

## 4.5 Constructed response examples

Table 8 presents constructed replies for the student-request route, not observed LLM outputs. Both groups have requested approved process help. The Brainstorming group has posted ideas for saving energy at school and is still generating; its entries do not mention a competition. The PPCo group selected window sealing earlier, has written concern questions that do not include cost, and has requested approved prioritisation help. These conditions allow each reply to be judged against work the groups actually retain.

**Table 8.** Constructed response pairs illustrating substitution in two techniques.

| Activity and judgement | Constructed candidate response | Why it does or does not fit |
|---|---|---|
| **Brainstorming: violation** | "Why not add a school-wide competition and make that your main direction?" | Supplies an idea and selects the group's focus during generation. |
| **Brainstorming: permissible** | "What different kinds of possibilities have you listed, and what else might your group explore?" | Invites the group to inspect and extend its own set without naming or ranking ideas. |
| **PPCo Concerns: violation** | "Shouldn't cost be the concern you tackle first, perhaps by applying for a grant?" | Supplies and prioritises a concern, then proposes a fix assigned to the group. |
| **PPCo Concerns: permissible** | "Which of your group's concern questions will you address first, and why?" | Invites the group to prioritise its own questions without choosing one for it. |

# 5. Discussion

The quality of a group's creative product cannot alone establish which work its members practised in making it (Aru, 2025; Green et al., 2024; Rhodes, 1961). This distinction matters when an LLM can introduce ideas and influence which ones a group pursues. The discussion follows the assignment of creative work from activity design (RQ1) through the proposed LLM workflow (RQ2), then considers its implications for group creativity.

## 5.1 From creative process to the work of one activity

RQ1 requires a way to specify AI support without treating a phase name as a complete description of learners' work. Creative-process models divide work differently, and groups may return to an earlier problem or idea (Botella et al., 2018; Lubart, 2001; Sawyer, 2021). The framework therefore locates the assignment in the current activity: the phase gives its purpose, the creative technique section specifies how learners work on an idea, and the task identifies what they are working on. A cue to broaden possibilities may serve Brainstorming generation but interrupt a PPCo step in which the group is appraising one selected idea. Rafner (2026) argues that research must describe how human–AI collaboration is organised rather than treat "AI use" as a single intervention. For creative learning, the activity specification records that organisation before support is given.

Accounts of an AI's role offer a complementary view. Lubart's (2005) *computer as coach* supports the person engaged in creative work; FERC organises human framing and judgement across repeated exchanges (Sherson et al., 2026); and Cho (2026) relates AI roles in mathematics learning to task design and teacher mediation. These accounts help explain the form of collaboration. A role label alone, however, cannot tell an educator whether a particular question supplies a Brainstorming idea or chooses a PPCo concern. The activity specification makes that judgement explicit, allowing support to change as the work assigned to learners changes.

That work includes control over the group's direction. Educational assistants may withhold direct answers while offering guidance (Kazemitabaar et al., 2024; Liffiton et al., 2023), but creative work can be displaced without a finished answer appearing in the reply. Supplying a concern learners were meant to formulate is *functional substitution*; choosing which concern deserves attention is *attentional substitution*. Creative metacognition distinguishes monitoring progress from controlling what to do next (Lebuda & Benedek, 2023), while socially shared regulation locates negotiation of direction within the group (Järvelä & Hadwin, 2013). A cue may draw attention without settling a decision: asking what members notice about the range of their own entries invites monitoring, whereas declaring that they have enough ideas closes generation for them. The relevant boundary is whether the LLM chooses a focus, criterion or next step that learners were assigned to decide.

Pedagogical objectives give these assignments their instructional purpose. During Brainstorming, broad participation and deferred criticism preserve time for possibilities to emerge (Osborn, 1953). While developing a provisional idea, a question about uncertainty may instead invite productive revision (Reiser, 2004). The educator decides which help serves the present activity and when it may appear. Creative projects can take unexpected directions (Sawyer, 2021), so an already approved activity can be selected when the group returns to it, while a new assignment requires review. Reusable specifications can reduce repeated preparation, but the quality and workload of educator decisions still need examination.

## 5.2 From educator judgement to constrained LLM behaviour

RQ2 concerns whether the approved assignment can be maintained during interaction. The four translation rules make the origin of each AI permission and prohibition inspectable: the support objective, role and timing, feedback scope and behavioural constraints each follow from the activity specification. Its machine-readable form, the activity contract, then governs the proposed workflow. Retrieved guidance informs a draft about the current activity but grants no additional permission. Student-authored text may attempt to redirect the model (Greshake et al., 2023), and a reply faithful to a retrieved record can still supply a concern or criterion belonging to the group. Source faithfulness (Es et al., 2024) and pedagogical fit therefore require different checks.

The response check is consequential because the same grammatical form can perform different work. A form check may establish that a draft uses an allowed structure; it cannot establish that a question leaves idea generation or priority setting to learners. The proposed functional judge assesses what the draft would contribute to the current activity. It could nevertheless accept a subtle substitution or reject useful help (Zheng et al., 2023). Human-labelled cases would be needed before classroom use to estimate both errors across the intended activities, languages and learners. A system that refuses every reply might avoid prohibited contributions but would provide little scaffolding. Usefulness must therefore be examined alongside compliance.

The value of permissible help also depends on who initiates it and when it arrives. Students may ask for process help; a request for an idea or decision assigned to the group does not transfer that work to the LLM. In addition, the system may offer a brief cue at an educator-approved checkpoint after the group has begun working with the technique. Offers require particular care because they can interrupt members as they negotiate an idea (Horvitz, 1999). Both routes follow the same learner-owned boundary, while students can defer or dismiss an offer. This arrangement gives learners access to help during the activity without making the LLM responsible for choosing the group's creative direction. A decline in offers would describe the controller's behaviour; growing learner independence (van de Pol et al., 2010) would have to be examined in unaided work.

### 5.3 What constrained support means for group creativity

The educational role of AI-generated content depends on what the group has been asked to do with it. An example can provide material for critique when critique is the assigned task. During idea generation, an example may supply an alternative students were meant to produce and influence what follows. Exposure to others' ideas can stimulate as well as constrain creative search (Dugosh et al., 2000; S. M. Smith et al., 1993). Findings on AI assistance also differ across tasks: AI story ideas improved ratings of individual stories while reducing diversity among them (Doshi & Hauser, 2024), whereas source-labelled primes had mostly null effects on divergent association scores (Medeiros et al., 2025). Neither study examined groups working through a structured creative technique. The framework adds a question for such learning: did students perform the creative work and decisions that the activity was intended to develop?

This question has a practical implication for project-based learning. A digital workspace that identifies the current phase, technique section and group entries could apply an approved specification as learners progress through their project. The educator would prepare specifications for activities in the project sequence using reusable technique records; the workspace would select the matching approved specification as the group moves between them. Compared with access to an unconfigured general-purpose LLM, this integration is intended to make support specific to the work assigned in each activity while reserving authorship and decisions for the group.

Whether the group retains those decisions depends on what happens after a reply is delivered. Members may adopt a cue, contest it, combine it with an earlier entry or ignore it. A permissible question may still interrupt a productive exchange or draw attention away from a quieter member's proposal. Accounts of distributed creativity describe ideas developing through interactions among people and shared representations (Glăveanu, 2014; Sawyer & DeZutter, 2009; Vallée-Tourangeau, 2026). Response checks can assess whether a draft meets the conditions for delivery. The educational value of this design depends on how students develop and direct ideas together after receiving help, and on what they can do later without AI.

## 6. Limitations and future research

The worked examples show how the framework applies to Viacocrea's phases, Plan Brainstorming and an illustrative PPCo activity in Build. Its application to other creative processes and techniques remains to be examined. Researchers can use the reusable forms to document their own phase and technique sources, then examine whether educators produce defensible activity specifications and how much preparation the procedure requires.

The constrained LLM workflow is specified in this paper and awaits integrated testing. Implementation studies need to establish whether prompts, retrieved guidance and response checks deliver useful support while detecting replies that perform learners' work or choose their direction. Testing should cover the intended activities, languages, age groups and both help routes. Use with minors also requires validated safety procedures and clear arrangements for processing student text. These studies can identify where the workflow follows the educator's specification and where its rules need revision.

Earlier Viacocrea studies measured novelty and flexibility in group-authored ideas and examined real-time novelty feedback (Ul Haq & Pifarré, 2026; Ul Haq et al., 2024a, 2024b). The present framework extends that work by specifying the creative operations and decisions students should retain when an LLM provides support. Future classroom studies can use the earlier measures alongside evidence from group discussion, revisions and participation, followed by unaided creative work. This progression connects the measurement of group ideas, the design of activity-specific AI support and the study of what learners do with that support.

## 7. Conclusion

Human–AI cocreation in project-based learning raises an educational question: what should students do and decide for themselves when an LLM can contribute throughout a creative project? A well-developed group product cannot establish who generated alternatives, appraised ideas or chose a direction. This paper argues that the work reserved for learners should be specified for each activity before AI support is introduced.

To address RQ1, the cognitive–pedagogical framework brings together the purpose of a creative-process phase, the instructions of the active creative-technique section and the current task. The educator records the cognitive and metacognitive objective, the pedagogical objective, the task and group artefact as context, and the product students must author. Four translation rules derive what help the LLM may give, when it may intervene and what its feedback may address. They also set limits against *functional substitution*, in which the LLM performs students' creative work, and *attentional substitution*, in which it chooses the group's focus, criteria or next step. To address RQ2, a proposed constrained workflow applies those decisions to student requests and feedback offers through prompting, approved retrieved guidance and checks before delivery. The LLM's role follows from the work assigned in the activity.

Educators can reuse and revise these specifications as a project develops. Creativity researchers can describe AI interventions in terms of the work assigned to learners, while educational and cognitive psychologists can examine how support affects idea generation, appraisal and shared control. For AI researchers, the specifications define requirements that a system can implement and test. These connections provide a basis for studying whether AI support advances creative learning while students retain agency over the ideas and decisions through which they learn.

## Declarations

**Materials availability.** Supplements S1–S3 accompany this preprint. Additional framework materials are available from Ijaz Ul Haq on reasonable request. The examples in Table 8 are constructed; no executable or validated system is provided.

**Ethics statement.** This conceptual paper collected no new participant data and did not reanalyse data from earlier studies. Future classroom research will require ethical approval, appropriate consent and safeguards for students' data, including data from minors.

**Acknowledgement and manuscript status.** Dr. Aamir Anwar contributed to the revision and further improvement of the manuscript. This preprint is a version of the manuscript currently under review.

**Competing interests.** The author declares no competing interests.

## References

Amabile, T. M. (1983). The social psychology of creativity: A componential conceptualization. *Journal of Personality and Social Psychology, 45*(2), 357–376. https://doi.org/10.1037/0022-3514.45.2.357

Anderson, B. R., Shah, J. H., & Kreminski, M. (2024). Homogenization effects of large language models on human creative ideation. In *Proceedings of the 16th Conference on Creativity & Cognition* (pp. 413–425). ACM. https://doi.org/10.1145/3635636.3656204

Aru, J. (2025). Artificial intelligence and the internal processes of creativity. *The Journal of Creative Behavior, 59*(2), e1530. https://doi.org/10.1002/jocb.1530

Bastani, H., Bastani, O., Sungu, A., Ge, H., Kabakcı, Ö., & Mariman, R. (2025). Generative AI without guardrails can harm learning: Evidence from high school mathematics. *Proceedings of the National Academy of Sciences, 122*(26), e2422633122. https://doi.org/10.1073/pnas.2422633122

Biggs, J. (1996). Enhancing teaching through constructive alignment. *Higher Education, 32*(3), 347–364. https://doi.org/10.1007/BF00138871

Botella, M., Zenasni, F., & Lubart, T. (2018). What are the stages of the creative process? What visual art students are saying. *Frontiers in Psychology, 9*, Article 2266. https://doi.org/10.3389/fpsyg.2018.02266

Celik, I., Zabolotna, K., & Viberg, O. (2026). Knowledge construction with GenAI: The role of theory-informed prompt engineering in achieving pedagogical alignment. *Innovations in Education and Teaching International*. Advance online publication. https://doi.org/10.1080/14703297.2026.2631586

Chi, M. T. H., & Wylie, R. (2014). The ICAP framework: Linking cognitive engagement to active learning outcomes. *Educational Psychologist, 49*(4), 219–243. https://doi.org/10.1080/00461520.2014.965823

Cho, H. (2026). Relocating the locus of generation: The AI-GLM framework for generative learning in AI-mediated mathematics education. *Educational Psychology Review*, *38*, Article 116. https://doi.org/10.1007/s10648-026-10210-y

Creative Education Foundation. (2015). *Creative problem solving resource guide* (Version 6.3). https://www.creativeeducationfoundation.org/wp-content/uploads/2015/06/CPS-Guide-6-3-web.pdf

Cropley, A. (2006). In praise of convergent thinking. *Creativity Research Journal, 18*(3), 391–404. https://doi.org/10.1207/s15326934crj1803_13

Diehl, M., & Stroebe, W. (1987). Productivity loss in brainstorming groups: Toward the solution of a riddle. *Journal of Personality and Social Psychology, 53*(3), 497–509. https://doi.org/10.1037/0022-3514.53.3.497

Doshi, A. R., & Hauser, O. P. (2024). Generative AI enhances individual creativity but reduces the collective diversity of novel content. *Science Advances, 10*(28), eadn5290. https://doi.org/10.1126/sciadv.adn5290

Dugosh, K. L., Paulus, P. B., Roland, E. J., & Yang, H.-C. (2000). Cognitive stimulation in brainstorming. *Journal of Personality and Social Psychology, 79*(5), 722–735. https://doi.org/10.1037/0022-3514.79.5.722

Es, S., James, J., Espinosa Anke, L., & Schockaert, S. (2024). RAGAs: Automated evaluation of retrieval augmented generation. In *Proceedings of the 18th Conference of the European Chapter of the Association for Computational Linguistics: System Demonstrations* (pp. 150–158). Association for Computational Linguistics. https://aclanthology.org/2024.eacl-demo.16/

Finke, R. A., Ward, T. B., & Smith, S. M. (1992). *Creative cognition: Theory, research, and applications*. MIT Press.

Getzels, J. W., & Csikszentmihalyi, M. (1976). *The creative vision: A longitudinal study of problem finding in art*. Wiley.

Glăveanu, V. P. (2014). *Distributed creativity: Thinking outside the box of the creative individual*. Springer. https://doi.org/10.1007/978-3-319-05434-6

Green, A. E., Beaty, R. E., Kenett, Y. N., & Kaufman, J. C. (2024). The process definition of creativity. *Creativity Research Journal, 36*(3), 544–572. https://doi.org/10.1080/10400419.2023.2254573

Greshake, K., Abdelnabi, S., Mishra, S., Endres, C., Holz, T., & Fritz, M. (2023). Not what you've signed up for: Compromising real-world LLM-integrated applications with indirect prompt injection. In *Proceedings of the 16th ACM Workshop on Artificial Intelligence and Security* (pp. 79–90). ACM. https://doi.org/10.1145/3605764.3623985

Guilford, J. P. (1967). *The nature of human intelligence*. McGraw-Hill.

Haatainen, O., & Aksela, M. (2021). Project-based learning in integrated science education: Active teachers' perceptions and practices. *LUMAT: International Journal on Math, Science and Technology Education, 9*(1), 149–173.

Hattie, J., & Timperley, H. (2007). The power of feedback. *Review of Educational Research, 77*(1), 81–112. https://doi.org/10.3102/003465430298487

Hines, K., Lopez, G., Hall, M., Zarfati, F., Zunger, Y., & Kiciman, E. (2024). *Defending against indirect prompt injection attacks with spotlighting* [Preprint]. arXiv. https://arxiv.org/abs/2403.14720

Horvitz, E. (1999). Principles of mixed-initiative user interfaces. In *Proceedings of the SIGCHI Conference on Human Factors in Computing Systems* (pp. 159–166). ACM. https://doi.org/10.1145/302979.303030

Isaksen, S. G., & Treffinger, D. J. (2004). Celebrating 50 years of reflective practice: Versions of creative problem solving. *The Journal of Creative Behavior, 38*(2), 75–101. https://doi.org/10.1002/j.2162-6057.2004.tb01234.x

Järvelä, S., & Hadwin, A. F. (2013). New frontiers: Regulating learning in CSCL. *Educational Psychologist, 48*(1), 25–39. https://doi.org/10.1080/00461520.2012.748006

Kazemitabaar, M., Ye, R., Wang, X., Henley, A. Z., Denny, P., Craig, M., & Grossman, T. (2024). CodeAid: Evaluating a classroom deployment of an LLM-based programming assistant that balances student and educator needs. In *Proceedings of the 2024 CHI Conference on Human Factors in Computing Systems* (pp. 1–20). ACM. https://doi.org/10.1145/3613904.3642773

Kumar, H., Vincentius, J., Jordan, E., & Anderson, A. (2025). Human creativity in the age of LLMs: Randomized experiments on divergent and convergent thinking. In *Proceedings of the 2025 CHI Conference on Human Factors in Computing Systems*. ACM. https://doi.org/10.1145/3706598.3714198

LearnLM Team. (2024). *LearnLM: Improving Gemini for learning* [Preprint]. arXiv. https://arxiv.org/abs/2412.16429

Lebuda, I., & Benedek, M. (2023). A systematic framework of creative metacognition. *Physics of Life Reviews, 46*, 161–181. https://doi.org/10.1016/j.plrev.2023.07.002

Lee, B. C., & Chung, J. J. (2024). An empirical investigation of the impact of ChatGPT on creativity. *Nature Human Behaviour, 8*, 1906–1914. https://doi.org/10.1038/s41562-024-01953-1

Lee, U., Shin, M., Jeong, Y., Lee, S., Moon, J., Joo, K., Lee, E., & Kwon, H. (2026). *LLMs are already good tutors: Training-free prompt optimization for pedagogical math tutoring* [Preprint]. arXiv. https://arxiv.org/abs/2605.27088

Lewis, P., Perez, E., Piktus, A., Petroni, F., Karpukhin, V., Goyal, N., Küttler, H., Lewis, M., Yih, W.-t., Rocktäschel, T., Riedel, S., & Kiela, D. (2020). Retrieval-augmented generation for knowledge-intensive NLP tasks. *Advances in Neural Information Processing Systems, 33*, 9459–9474. https://proceedings.neurips.cc/paper/2020/hash/6b493230205f780e1bc26945df7481e5-Abstract.html

Liffiton, M., Sheese, B., Savelka, J., & Denny, P. (2023). CodeHelp: Using large language models with guardrails for scalable support in programming classes. In *Proceedings of the 23rd Koli Calling International Conference on Computing Education Research* (pp. 1–11). ACM. https://doi.org/10.1145/3631802.3631830

Lubart, T. I. (2001). Models of the creative process: Past, present and future. *Creativity Research Journal, 13*(3–4), 295–308. https://doi.org/10.1207/S15326934CRJ1334_07

Lubart, T. I. (2005). How can computers be partners in the creative process: Classification and commentary on the special issue. *International Journal of Human-Computer Studies*, *63*(4–5), 365–369. https://doi.org/10.1016/j.ijhcs.2005.04.002

Medeiros, K., Cropley, D. H., Marrone, R. L., & Reiter-Palmon, R. (2025). Human–AI co-creativity: Does ChatGPT make us more creative? *The Journal of Creative Behavior, 59*(2), e70022. https://doi.org/10.1002/jocb.70022

Meincke, L., Nave, G., & Terwiesch, C. (2025). ChatGPT decreases idea diversity in brainstorming. *Nature Human Behaviour, 9*(6), 1107–1109. https://doi.org/10.1038/s41562-025-02173-x

Mumford, M. D., Mobley, M. I., Uhlman, C. E., Reiter-Palmon, R., & Doares, L. M. (1991). Process analytic models of creative capacities. *Creativity Research Journal, 4*(2), 91–122. https://doi.org/10.1080/10400419109534380

Nijstad, B. A., & Stroebe, W. (2006). How the group affects the mind: A cognitive model of idea generation in groups. *Personality and Social Psychology Review, 10*(3), 186–213. https://doi.org/10.1207/s15327957pspr1003_1

OECD. (2023). *PISA 2022 assessment and analytical framework*. OECD Publishing. https://doi.org/10.1787/dfe0bf9c-en

Osborn, A. F. (1953). *Applied imagination: Principles and procedures of creative problem solving*. Scribner.

Parasuraman, R., Sheridan, T. B., & Wickens, C. D. (2000). A model for types and levels of human interaction with automation. *IEEE Transactions on Systems, Man, and Cybernetics—Part A: Systems and Humans, 30*(3), 286–297. https://doi.org/10.1109/3468.844354

Paulus, P. B., & Brown, V. R. (2007). Toward more creative and innovative group idea generation: A cognitive-social-motivational perspective of brainstorming. *Social and Personality Psychology Compass, 1*(1), 248–265. https://doi.org/10.1111/j.1751-9004.2007.00006.x

Paulus, P. B., & Yang, H.-C. (2000). Idea generation in groups: A basis for creativity in organizations. *Organizational Behavior and Human Decision Processes, 82*(1), 76–87. https://doi.org/10.1006/obhd.2000.2888

Pifarré, M. (2023). Designing, implementing and evaluating a co-creative support technology. In *EDULEARN23 Proceedings* (pp. 4364–4367). IATED. https://doi.org/10.21125/edulearn.2023.1146

Puccio, G. J., Mance, M., & Murdock, M. C. (2011). *Creative leadership: Skills that drive change* (2nd ed.). Sage.

Rafner, J. (2026). Human–AI collaboration is an underspecified intervention: A reporting guide for cumulative research. *Creativity Research Journal*. Advance online publication. https://doi.org/10.1080/10400419.2026.2726243

Rebedea, T., Dinu, R., Sreedhar, M. N., Parisien, C., & Cohen, J. (2023). NeMo Guardrails: A toolkit for controllable and safe LLM applications with programmable rails. In *Proceedings of the 2023 Conference on Empirical Methods in Natural Language Processing: System Demonstrations* (pp. 431–445). Association for Computational Linguistics. https://doi.org/10.18653/v1/2023.emnlp-demo.40

Reiser, B. J. (2004). Scaffolding complex learning: The mechanisms of structuring and problematizing student work. *Journal of the Learning Sciences, 13*(3), 273–304. https://doi.org/10.1207/s15327809jls1303_2

Reiter-Palmon, R., Mumford, M. D., O'Connor Boes, J., & Runco, M. A. (1997). Problem construction and creativity: The role of ability, cue consistency, and active processing. *Creativity Research Journal, 10*(1), 9–23. https://doi.org/10.1207/s15326934crj1001_2

Reiter-Palmon, R., & Robinson, E. J. (2009). Problem identification and construction: What do we know, what is the future? *Psychology of Aesthetics, Creativity, and the Arts, 3*(1), 43–47. https://doi.org/10.1037/a0014629

Rezwana, J., & Maher, M. L. (2023). Designing creative AI partners with COFI: A framework for modeling interaction in human–AI co-creative systems. *ACM Transactions on Computer-Human Interaction, 30*(5), Article 67. https://doi.org/10.1145/3519026

Rhodes, M. (1961). An analysis of creativity. *Phi Delta Kappan, 42*(7), 305–310.

Rietzschel, E. F., Nijstad, B. A., & Stroebe, W. (2006). Productivity is not enough: A comparison of interactive and nominal groups in idea generation and selection. *Journal of Experimental Social Psychology, 42*(2), 244–251. https://doi.org/10.1016/j.jesp.2005.04.005

Rietzschel, E. F., Nijstad, B. A., & Stroebe, W. (2010). The selection of creative ideas after individual idea generation: Choosing between creativity and impact. *British Journal of Psychology, 101*(1), 47–68. https://doi.org/10.1348/000712609X414204

Risko, E. F., & Gilbert, S. J. (2016). Cognitive offloading. *Trends in Cognitive Sciences, 20*(9), 676–688. https://doi.org/10.1016/j.tics.2016.07.002

Runco, M. A., & Jaeger, G. J. (2012). The standard definition of creativity. *Creativity Research Journal, 24*(1), 92–96. https://doi.org/10.1080/10400419.2012.650092

Sawyer, R. K. (2012). *Explaining creativity: The science of human innovation* (2nd ed.). Oxford University Press.

Sawyer, R. K. (2021). The iterative and improvisational nature of the creative process. *Journal of Creativity, 31*, 100002. https://doi.org/10.1016/j.yjoc.2021.100002

Sawyer, R. K. (2022). The dialogue of creativity: Teaching the creative process by animating student work as a collaborating creative agent. *Cognition and Instruction, 40*(4), 459–487. https://doi.org/10.1080/07370008.2021.1958219

Sawyer, R. K., & DeZutter, S. (2009). Distributed creativity: How collective creations emerge from collaboration. *Psychology of Aesthetics, Creativity, and the Arts, 3*(2), 81–92. https://doi.org/10.1037/a0013282

Sherson, J., Rafner, J., Reiter-Palmon, R., Lebuda, I., Brosbøl Kjeldsen, F., Söllner, M., Kenett, Y., Emruli, B., Rindle, M., Weiss, S., Goecke, B., Vinchon, F., Mao, Y., Dellermann, D., Rahimi, S., Marler, J. H., Nguyen, A., Bjerring, J. C., & DiPaola, S. (2026). Ensuring human authorship in the age of generative AI: The FERC framework for governed human–AI collaboration. In *NordiCHI '26 Adjunct: Adjunct Proceedings of the 14th Nordic Conference on Human-Computer Interaction* (pp. 1–11). ACM. https://doi.org/10.1145/3821402.3830141

Smith, G. F. (1998). Idea-generation techniques: A formulary of active ingredients. *The Journal of Creative Behavior, 32*(2), 107–134. https://doi.org/10.1002/j.2162-6057.1998.tb00810.x

Smith, S. M., Ward, T. B., & Schumacher, J. S. (1993). Constraining effects of examples in a creative generation task. *Memory & Cognition, 21*(6), 837–845. https://doi.org/10.3758/BF03202751

Sonkar, S., Ni, K., Chaudhary, S., & Baraniuk, R. (2024). Pedagogical alignment of large language models. In *Findings of the Association for Computational Linguistics: EMNLP 2024* (pp. 13641–13650). Association for Computational Linguistics. https://doi.org/10.18653/v1/2024.findings-emnlp.797

Sun, M., Wang, M., Wegerif, R., & Peng, J. (2022). How do students generate ideas together in scientific creativity tasks through computer-based mind mapping? *Computers & Education, 176*, 104359. https://doi.org/10.1016/j.compedu.2021.104359

Ul Haq, I., & Pifarré, M. (2026). Promoting cocreativity with real-time automatic novelty feedback in open-ended problem-solving. *Journal of Computers in Education*. Advance online publication. https://doi.org/10.1007/s40692-026-00390-5

Ul Haq, I., Pifarré, M., & Fraca, E. (2024a). Natural language processing approach to evaluate real-time flexibility of ideas to support collaborative creative process. *International Journal of Emerging Technologies in Learning, 19*(5), 93–107. https://doi.org/10.3991/ijet.v19i05.47465

Ul Haq, I., Pifarré, M., & Fraca, E. (2024b). Novelty evaluation using sentence embedding models in open-ended cocreative problem-solving. *International Journal of Artificial Intelligence in Education, 34*, 1599–1626. https://doi.org/10.1007/s40593-024-00392-3

Vallée-Tourangeau, F. (2026). The one-I model of creativity: A commentary on Green et al. (2024). *Creativity Research Journal, 38*(3), 684–689. https://doi.org/10.1080/10400419.2025.2534806

van de Pol, J., Volman, M., & Beishuizen, J. (2010). Scaffolding in teacher–student interaction: A decade of research. *Educational Psychology Review, 22*, 271–296. https://doi.org/10.1007/s10648-010-9127-6

Viacocrea. (n.d.). *El método Cocrea: Objetivos de las fases*. https://viacocrea.eu/index.php/pedagogia/

Vinchon, F., Lubart, T., Bartolotta, S., Gironnay, V., Botella, M., Bourgeois-Bougrine, S., Burkhardt, J.-M., Bonnardel, N., Corazza, G. E., Glăveanu, V., Hanson, M. H., Ivcevic, Z., Karwowski, M., Kaufman, J. C., Okada, T., Reiter-Palmon, R., & Gaggioli, A. (2023). Artificial intelligence & creativity: A manifesto for collaborative human creativity. *The Journal of Creative Behavior, 57*(4), 472–484. https://doi.org/10.1002/jocb.597

Wallace, E., Xiao, K., Leike, R., Weng, L., Heidecke, J., & Beutel, A. (2024). *The instruction hierarchy: Training LLMs to prioritize privileged instructions* [Preprint]. arXiv. https://arxiv.org/abs/2404.13208

Wallas, G. (1926). *The art of thought*. Harcourt, Brace.

White, J., Fu, Q., Hays, S., Sandborn, M., Olea, C., Gilbert, H., Elnashar, A., Spencer-Smith, J., & Schmidt, D. C. (2023). *A prompt pattern catalog to enhance prompt engineering with ChatGPT* [Preprint]. arXiv. https://arxiv.org/abs/2302.11382

Zheng, L., Chiang, W.-L., Sheng, Y., Zhuang, S., Wu, Z., Zhuang, Y., Lin, Z., Li, Z., Li, D., Xing, E. P., Zhang, H., Gonzalez, J. E., & Stoica, I. (2023). Judging LLM-as-a-judge with MT-Bench and Chatbot Arena. *Advances in Neural Information Processing Systems, 36,* 46595–46623. https://papers.neurips.cc/paper_files/paper/2023/hash/91f18a1287b398d378ef22505bf41832-Abstract-Datasets_and_Benchmarks.html